\documentclass[a4paper]{article}
\usepackage{natbib}
\usepackage[margin=1in]{geometry}
\usepackage{amsmath}
\usepackage{amssymb}
\usepackage{bm}
\usepackage{graphicx}
\usepackage[linesnumbered,lined,boxruled,algoruled]{algorithm2e}
\usepackage{lineno}

\newcommand{\mb}{\mathbf}
\newcommand{\T}{^\mathrm T}
\newcommand{\an}{^\mathrm{a}}
\newcommand{\fc}{^\mathrm{f}}

\title{\textbf{On building the state error covariance from a state estimate}}
\author{Pavel Sakov\\ Bureau of Meteorology, Australia}
\date{\today}

\begin{document}

\maketitle

\begin{abstract}
  It was recently found with the aid of machine learning that for a variety of toy data assimilation systems with chaotic Lorenz-96 model it is possible to achieve a nearly-optimal data assimilation without carrying the state error covariance between cycles.
  This result does not look surprising on its own because not carrying covariance is the approach taken by standard 4D-Var, but it was found ``astonishing'' in the context of the machine learning-based system trained on the ensemble Kalman filter.
  This note proposes two algorithms for building the state error covariance from a state estimate that yield good performance and could be worked out by the deep learning-based system.
\end{abstract}

\section{Introduction}

A recent study by \citet{boc24b} presents results of application of deep learning to discovering data assimilation (DA) schemes.
They found that ``the accuracy of the states obtained from the learned analysis approaches that of the best possibly tuned ensemble Kalman filter, and is far better than that of variational DA alternatives''.
Moreover, it turned out that ``this can be achieved while propagating even just a single state in the forecast step''.

To give some perspective on why this result is unexpected we note that the DA methods used with large-scale models are essentially linear.
Consistent formulations of these methods involve characterising the state of the DA system (SDAS) by a set of two objects: the model state $\mb x$ and the model state error covariance $\mb P$; the latter is necessary for obtaining the sensitivities of the analysis increments to innovations (differences between observations and state).
The SDAS is used to assimilate new information (observations); is updated during assimilation; and needs either to be propagated in time between (and, possibly, within) data assimilation cycles or built at each cycle.

Evolving the covariance is, typically, by far the most computationally expensive part of a DA cycle for Kalman filter (KF) based systems.
The fact that a deep-learning based DA system (DLS) was able to match a well tuned ensemble Kalman filter (EnKF) without propagating the state error covariance between cycle means that it is able to build it well enough from a state estimate.

Upon consideration, this fact in itself is not unexpected, because this is an approach taken by 4D-Var, which allows it to basically match the performance of the EnKF in e.g. standard test with the Lorenz-96 model \citep[L96;][]{lor98a}.

Here by the ``standard test'' we consider the one described in \citet{whi02a}: perfect model framework; time steps of 0.05; integration with the 4-order Runger-Kutta method; observing every element of the state at each time step; Gaussian observation error with variance of 1.
(We will refer to this test hereafter as Standard Test, or ST.)
A well tuned EnKF can reach average filtering analysis RMSE of about 0.179.
According to \citet[][fig. 3]{boc13a}, that can be matched by 4D-Var with data assimilation window of about 31 (or about 4 average error doubling times).
Therefore, it \emph{is} possible to build a good enough estimate of the state error covariance from the state estimate without carrying it between cycles.
Having said that, we need to note that, firstly, 4D-Var is an iterative method and as such should be compared performance-wise with the \emph{iterative} ensemble Kalman filter (IEnKF); and secondly, using long data assimilation window means using the corresponding set of past observations, which is not the case for the DLS in \citet{boc24b}.
At the same time, systems with static covariance show much inferior filtering performance: ensemble optimal interpolation (EnOI) systems with static covariance derived from a long model run can achieve RMSE of about 0.41; and 3D-Var systems are reported to achieve RMSE of 0.40 \citep{boc24b}.

In this context, it was unexpected when \citet{boc24b} found that a DLS trained on the EnKF achieved analysis RMSE of 0.185 propagating just a single state, almost matching that of the best tuned EnKF.
Below we consider two algorithms for building the state error covariance based on the state estimate that would allow to achieve this result.

The outline of this study is as follows.
Section~\ref{sec:preliminary} provides some preliminary considerations on the problem; Section~\ref{sec:twoapproaches} proposes two algorithms for building the state error covariance from the state estimate; the results are discussed Section~\ref{sec:discussion}; and the summary is given in Section~\ref{sec:summary}.

\section{Preliminary considerations}
\label{sec:preliminary}

To the best of our knowledge, all known rigorous DA methods involve using estimated state error covariance, explicitly or implicitly.
(Including the particle filters, which attempt estimating the whole probability distribution function.)
We can therefore assume that the deep learning-based system in \citet{boc24b} (referred to as the analysis operator $a_\theta$) is able to reconstruct the dynamic covariance well enough to perform almost equally with the EnKF.

Note that the estimates of the state error covariance in a spun-up chaotic system are largely based on information from the past observations (with the aid of the model, which is known to $a_\theta$).
This means that if the past observations were made available then the state error covariance could be rebuilt from scratch at each cycle by re-running the system from some prior time.
However, the trained DA operator $a_\theta$ is completely static; in other words, given the current state estimate and observations it is supposed to yield the same increment regardless of the history of the system.

In \citet{boc24b} $a_\theta$ was formulated in the EnKF context as a function of the forecast ensemble and observations:
\begin{align}
  \label{atheta}
  \mb E_k\an = \mb E_k\fc + a_\theta(\mb E_k\fc, \mb H_k\T \mb R_k^{-1} \delta \mb y_k),
\end{align}
where $k$ is the cycle number, $\mb E_k\an$ is the analysis ensemble, $\mb E_k\fc$ -- forecast ensemble, $a_\theta$ -- trained analysis operator, the term $\mb H_k\T \mb R_k^{-1} \delta \mb y_k$ represents standardised (scaled) innovations, $\mb H_k\T$ -- linearised observation operator, $\mb R_k$ -- observation error covariance, and $\delta \mb y_k$ -- innovations.
By design, the EnKF ensemble carries both the state and covariance estimates.
The discussed result refers to the case when the ensemble is shrunk to a single member, i.e. no longer carries the state error covariance estimate between cycles:
\begin{align}
  \label{single}
  \mb x_k\an = \mb x_k\fc + a_\theta(\mb x_k\fc, \mb H_k\T \mb R_k^{-1} \delta \mb y_k).
\end{align}
The fact that the trained $a_\theta$ in (\ref{single}) can perform nearly as well as that in (\ref{atheta}) with a large ensemble means that effectively it is able to build the state error covariance estimate $\mb P_k\fc$ from the state estimate $\mb x_k\fc$ and observations $\{\mb y_k, \mb H_k, \mb R_k\}$, and that the built covariance is of similar quality to that in a well tuned EnKF.

As suggested by the functional form of $a_\theta$ in (\ref{single}), potentially the covariance could be built based on both the state estimate and current observations.
However, the observations in ST (in a properly set spun-up system) can be characterised as ``noisy'': their standard error deviation is equal to 1, while the standard deviation of state estimate is below 0.2.
Therefore, it is essentially impossible to get much information about the state error from innovations.
This means that for ST it is possible to build a good estimate of the state error covariance based on the state estimate only:
\begin{align}
  \label{xonly}
  {a_\theta}_k = a_\theta(\mb x_k\fc).
\end{align}

\section{Building covariance from the state: two algorithms}
\label{sec:twoapproaches}

Armed with this information, we can try to find a way to build an estimate of the state error covariance based on the state estimate only:
\begin{align}
  \mb P_k\fc = \mb P(\mb x_k\fc)
\end{align}
such that when used in a KF based system would yield RMSE in the ballpark of a well tuned EnKF (0.18) rather than of EnOI or 3D-Var (0.40).

In the EnKF the state error covariance is handled implicitly via ensemble anomalies $\mb A$: $\mb P = \mb A \mb A\T / (m - 1)$, where $m$ is the ensemble size.
Below we will consider ``ensemble anomalies'' in this context, i.e. as an equivalent representation of the covariance.

We first note that a ``normally'' functioning DA system implies that the model is ``linearly constrained''.
By that we mean that the estimated model state is close enough to the true state so that the the evolution of anomalies of magnitude about the uncertainty of the state estimate are mainly governed by the tangent linear model (TLM) to the true (or estimated) trajectory.
Such a state of a DA system can also be characterised as ``weakly nonlinear''.

As a consequence, the ensemble anomalies in a weakly nonlinear EnKF system should mainly belong to the tangent linear space of the model and evolve with the TLM.
(``Mainly'' because generally, due to model nonlinearity, ensemble members are normally ``slightly'' off the model attractor.)
At a given time the ensemble anomalies in a weakly nonlinear EnKF system with chaotic model can be seen as being in balance between growth caused by model instabilities and dumping by DA corrections.
As a crude approximation one can neglect the impact of DA and assume that the modal composition of the covariance is mainly defined by the model dynamics over a relatively short time interval prior to the analysis.
This is implemented in Algorithm~1.

\vspace{0.3cm}
\begin{algorithm}[H]
  \DontPrintSemicolon
  \SetKwInput{KwInput}{Input}
  \SetKwInput{KwOutput}{Output}
  \KwInput{$\mb x\fc \in R^n$ -- state estimate}
  \KwInput{$\mb M_k$ -- propagator of state $\mb x\fc$ by $k$ timesteps: $\mb x_k = \mb M_k(\mb x\fc)$}
  \KwInput{$\mb L_k$ -- tangent linear propagator of perturbation $\mb y_0$ over trajectory from $\mb x$ by $k$ timesteps: $\mb y_k = \mb L_k(\mb x, \mb y_0)$}
  \KwInput{$T$ -- number of time steps (parameter)}
  \KwInput{$\varepsilon$ -- perturbation amplitude (parameter)}
  \KwOutput{$\mb P\fc$ -- state error covariance}
  $\mb x_{-T} \leftarrow \mb M_{-T}(\mb x\fc)$\;
  $\mb A \leftarrow \mb L_T(\mb x_{-T}, \varepsilon \mb I)$\;
  $\mb P\fc \leftarrow \mb A \mb A\T / n$\;
  \caption{Generating state error covariance from state estimate}
\end{algorithm}
\vspace{0.3cm}

Algorithm~1 (A1) first propagates the forecast state back by $T$ timesteps.
Then it propagates with the TLM a full-rank ensemble of perturbations (represented by the scaled identity matrix $\varepsilon \mb I$ at line 2 of A1) forward to the analysis time.
The propagated ensemble of perturbations is used in the analysis as a factorisation of the state error covariance.
The number of timesteps $T$ and perturbation amplitude $\varepsilon$ are the two tuneable parameters of the algorithm.

Using A1 for generating covariance in ST with algorithm parameters $T = 6$ and $\varepsilon = 0.925$ yields analysis RMSE of 0.235.
(Averaged over 10,000 cycles after a 400 cycles spin-up; a tuned EnKF run for the same true trajectory and observations yields RMSE of 0.180).
This performance is worse than that of the EnKF, but clearly is closer to the ballpark of the EnKF than of EnOI or 3D-Var.

\begin{figure}[h!]
  \centering
  \includegraphics{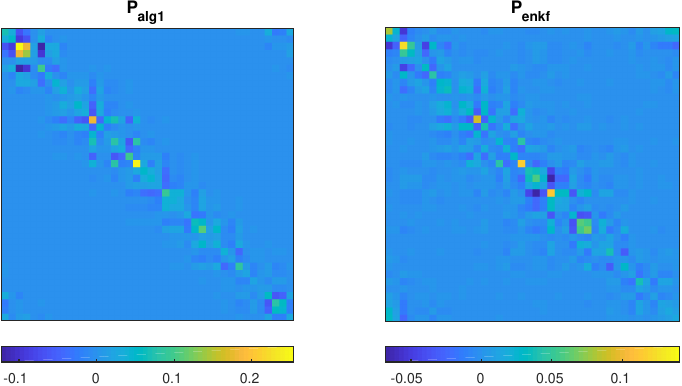}
  \caption{Comparison of covariances from Algorithm~1 and from the EnKF at a random cycle. The displayed range on the right plot is that from the left plot scaled by the ratio of the traces of the covariances.}
  \label{fig:p_alg1}
\end{figure}

\begin{figure}[h!]
  \centering
  \includegraphics[width=12cm]{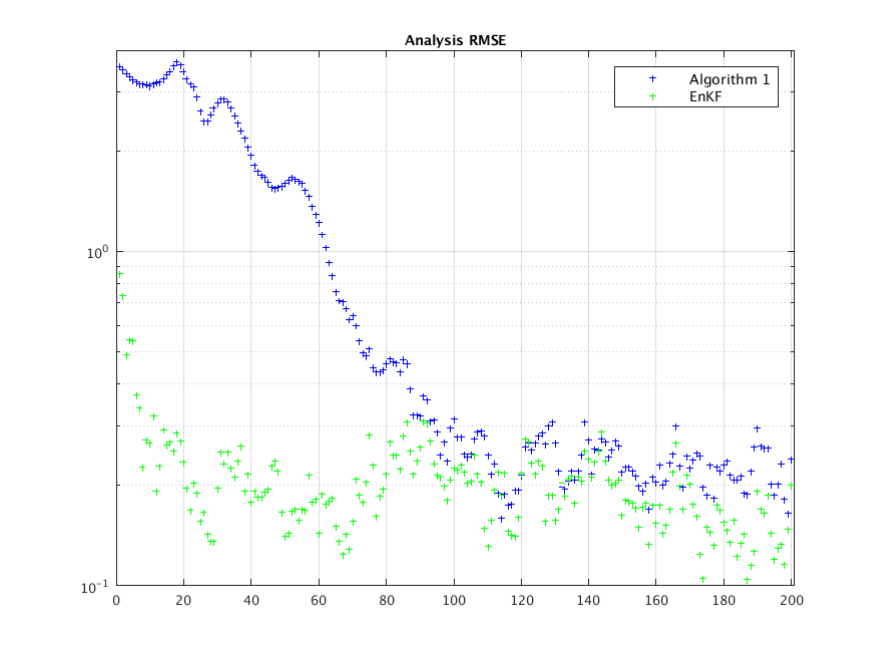}
  \caption{Analysis RMSE from the system with Algorithm 1 and from the EnKF during spin-up.}
  \label{fig:spinup1}
\end{figure}

Figure~\ref{fig:p_alg1} compares state error covariances from A1 ($\mb P_\mathrm{alg1}$) and from the EnKF ($\mb P_\mathrm{enkf}$) at a random cycle.
Although they have structural similarity, there are also substantial visual differences.
Interestingly, $\mb P_\mathrm{alg1}$ is naturally localised, showing basically zero covariances between distant elements.
This is due to the spatially limited propagation of initially localised perturbations by the tangent linear model in the interval of 6 timesteps used by the algorithm.
Another distinction of the system with covariance generated by A1 from the EnKF is a much slower spin-up shown in Figure~\ref{fig:spinup1}.
This is caused by the non-adaptive nature of A1, which is tuned to maximise performance of a spun system, and can be improved by using initially larger perturbation amplitudes.


\newpage
Now we will improve A1 by addressing the following considerations.
Firstly, while the internal spin-up time interval $T = 6$ maximises the performance of A1, it is obviously too short for the best performance, even shorter than the average error growth doubling time for L96, which is about 8 time steps.
However, increasing it leads to an excessive growth of some perturbations because A1 does not restrict it.
In the EnKF the growth of ensemble anomalies caused by model instabilities is balanced by the reduction of the estimated uncertainty in the analysis.
Implementing this reduction leads us to Algorithm~2 (A2), which damps the perturbations at each cycle by mimicking the ensemble anomalies update in the EnKF.

\begin{algorithm}[H]
  \DontPrintSemicolon
  \SetKwInput{KwInput}{Input}
  \SetKwInput{KwOutput}{Output}
  \KwInput{$\mb x\fc \in R^n$ -- state estimate}
  \KwInput{$\mb M_k$ -- propagator of state $\mb x\fc$ by $k$ timesteps: $\mb x_k = \mb M_k(\mb x\fc)$}
  \KwInput{$\mb L_k$ -- tangent linear propagator of perturbation $\mb y_0$ over trajectory from $\mb x$ by $k$ timesteps: $\mb y_k = \mb L_k(\mb x, \mb y_0)$}
  \KwInput{$\mb H_k$ -- observation operator at $k$ timesteps from analysis}
  \KwInput{$\mb R_k$ -- observation error covariance at $k$ timesteps from analysis}
  \KwInput{$T$ -- number of time steps (parameter)}
  \KwInput{$\varepsilon$ -- perturbation amplitude (parameter)}
  \KwOutput{$\mb P\fc$ -- state error covariance}
  $\mb x_{-T} \leftarrow \mb M_{-T}(\mb x\fc)$\;
  $\mb A = \varepsilon \mb I$\;
  \For {$k = -T \dots -1$}{
    $\mb A \leftarrow \mb L_1(\mb x_k, \mb A)$\;
    $\mb S \leftarrow \mb H_k \mb A \mb R_k^{-1/2} / n^{1/2}$\;
    $\mb A \leftarrow \mb A (\mb I + \mb S\T \mb S)^{-1/2}$\;
    $\mb x_{k+1} \leftarrow \mb M_1(\mb x_k)$\;
  }
  $\mb P\fc \leftarrow \mb A \mb A\T / n$\;
  \caption{Generating state error covariance from state estimate}
\end{algorithm}
\vspace{0.3cm}

Rerunning the test conducted earlier with A1 with A2, with parameters $T = 25$, $\varepsilon = 0.8$, yields analysis RMSE of 0.181, basically matching the EnKF.
Comparison of the generated covariance at the same cycle as in Figure~\ref{fig:p_alg1} with the EnKF covariance shows very good correspondence between the two.

The damping of perturbations in A2 makes it possible to use much longer time interval (of about 3 doubling times) for generating the covariance.
The approximately 4 times longer internal time interval and the internal perturbation damping make A2 substantially slower than A1.

\begin{figure}[h!]
  \centering
  \includegraphics{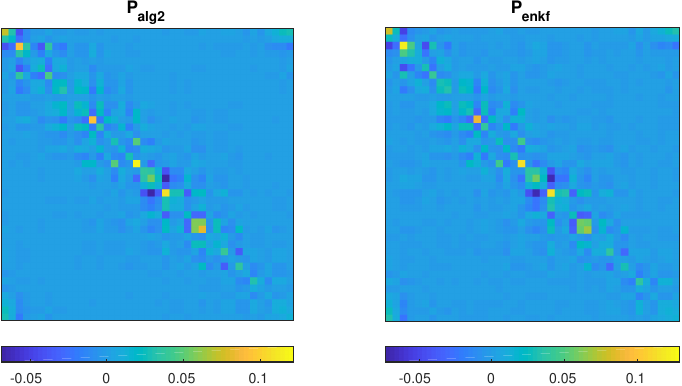}
  \caption{Comparison of covariances from Algorithm~2 and from the EnKF at the same cycle as in Fig.~\ref{fig:p_alg1}. The displayed range on the right plot is that from the left plot scaled by the ratio of the traces of the covariances.}
  \label{fig:p_alg2}
\end{figure}

\section{Discussion}
\label{sec:discussion}

The previous section proposed two algorithms for building the state error covariance from a state estimate.
Similar to the DLS operator $a_\theta$ in \citet{boc24b}, these algorithms are completely static in the sense that they depend on the state estimate only and do not depend on the previous evolution of the system.
Using these algorithms to build covariances in the ST makes it possible to reach performance in the approximate range of a well tuned EnKF.

The main purpose of these algorithms is to demonstrate that in the ST it is possible to obtain the DA performance of a well tuned EnKF without propagating the state error covariance; therefore, that achievement of DLS in \citet{boc24b} may not look as ``astonishing'' as it might have initially.

Note that the both algorithms use the TLM of L96.
It is not known whether the TLM was available for the DLS; however, the availability of the TLM for the algorithms should not be considered as an unfair advantage because once the forward model is available the TLM also becomes available via a few linear operations.

Both algorithms need to be tuned to the DA conditions.
To a large degree this tuning is possible due to the stationary nature of the ST: the number of observations, their positions and error variance, as well as the model settings (the forcing parameter) are constant in time.
In non-stationary conditions such a tuning would not be possible.
We speculate that a DLS would be presented with a much more difficult challenge in that case.

Having said that, we note that both algorithms, particularly A2, would be able to accommodate some general information about the past conditions (such as the observation operator and observation error covariance).

Concerning the computational performance of the algorithms, although their aim was to demonstrate the possibility of building the state error covariance from the state estimate, they are obviously much more computationally expensive than the EnKF.
To get a covariance estimate they require propagation of a full-rank ensemble of perturbations over multiple cycles, which is in a stark contrast with the EnKF, which represents a very economical framework for carrying and updating the SDAS.

Given the crudeness of A1, its level of performance looks rather surprising to us.
In particular, it indicates that a 4D-Var system can build a fairly good approximation of the flow dependent state error covariance using DA window of just about the error doubling time of the model.
It is not clear though to what degree this conclusion is specific to L96.

\begin{figure}[h!]
  \centering
  \includegraphics[width=12cm]{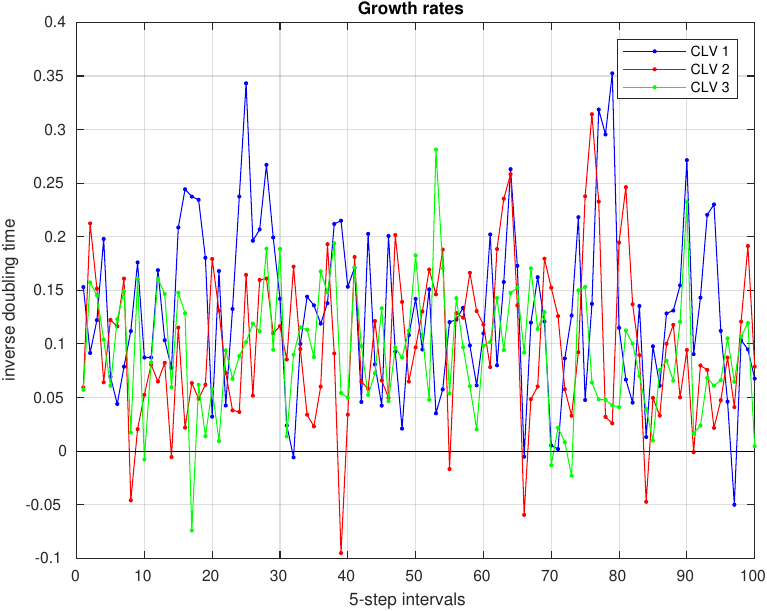}
  \caption{Growth rates of the first 3 CLVs of L96 for a random sequence of 100 5-timestep intervals.}
  \label{fig:clv_rates}
\end{figure}

\begin{figure}[h!]
  \centering
  \includegraphics[width=12cm]{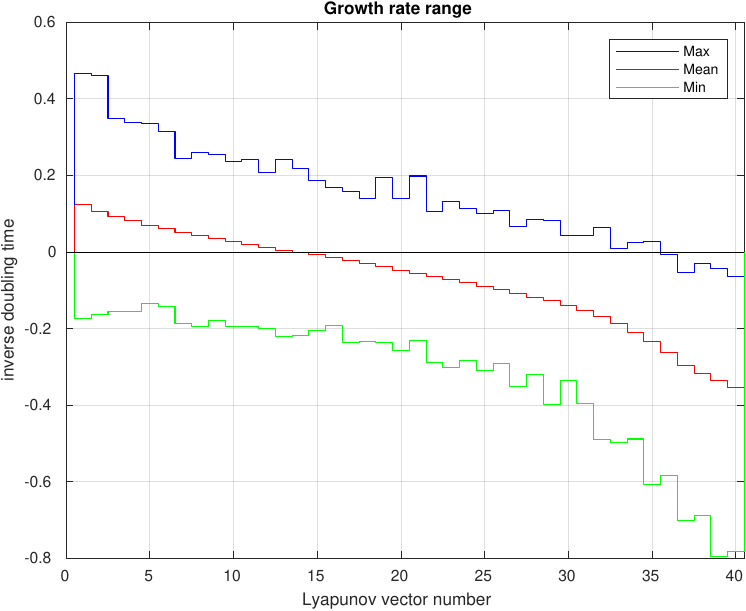}
  \caption{Minimal, mean, and maximal growth rates of CLVs of L96 during a random sequence of 4000 5-timestep intervals.}
  \label{fig:clv_rates_minmax}
\end{figure}

Figure~\ref{fig:clv_rates} shows the instantaneous growth rates for the first three covariant Lyapunov vectors (CLVs) (ordered by the time averaged growth rate) calculated for 100 consecutive 5-timestep intervals.\footnote{It is probably more precise to attribute these instantaneous growth rates to forward Gram-Schmidt vectors representing orthogonalised CLVs (see e.g. \citealt{gin13a}, sec.~3.2).}
It follows that the growth rates of these CLVs can change dramatically in just 5 timesteps.
Such variability is a common behaviour for all CLVs of L96, which is demonstrated in Figure~\ref{fig:clv_rates_minmax} showing maximal, mean and minimal growth rates for each CLV calculated for 4,000 consecutive 5-timestep intervals.
(Note that the fastest rates recorded correspond to the perturbation doubling time of just under 2 timesteps!)
Overall, the pattern of instabilities in L96 can be considered as a sequence of ``bursts'' of certain CLVs over rather short time intervals.
This behaviour may contribute to the decent performance of A1 with such a short internal spin-up interval ($T = 6$).

In contrast to A1, A2 exploits the fact that the evolution of covariance in the Kalman filter depends on observation locations and error variances but is mainly independent from innovations.
This makes possible an accurate (albeit computationally expensive) reconstruction of the covariance from the state estimate in weakly nonlinear systems.\footnote{We also note a remarkable similarity observed in state error covariance structure (such as exposed by the SST ensemble spread at https://github.com/sakov/enkf-c) from very different ocean forecasting systems regardless of the specific model and system settings.}

\section{Summary}
\label{sec:summary}

This note proposes two algorithms for building the state error covariance from a state estimate.
In KF-like systems using the covariance from these algorithms, in the ``standard'' DA test with L96 Algorithm~1 achieves analysis RMSE of 0.235, while Algorithm~2 achieves RMSE of 0.181 versus 0.180 by a well tuned EnKF.
Both algorithms are completely static in the sense that they depend on the state estimate only and do not depend on the previous evolution of the system.
Therefore, they could be potentially worked out by a deep learning-based system when conducting a run with a single ensemble member only.
The algorithms exploit the stationarity of ST.
We speculate that it would be much harder or even impossible for a DLS to match the EnKF performance in non-stationary conditions.

\clearpage
\bibliographystyle{ametsoc}
\bibliography{refs}

\begin{thebibliography}{5}
\expandafter\ifx\csname natexlab\endcsname\relax\def\natexlab#1{#1}\fi
\expandafter\ifx\csname url\endcsname\relax
  \def\url#1{{\tt #1}}\fi
\expandafter\ifx\csname urlprefix\endcsname\relax\def\urlprefix{URL }\fi
\expandafter\ifx\csname doiprefix\endcsname\relax\def\doiprefix{doi:}\fi

\bibitem[{Bocquet et~al.(2024)Bocquet, Farchi, Finn, Durand, Cheng, Chen,
  Pasmans, and Carrassi}]{boc24b}
Bocquet, M., A.~Farchi, T.~S. Finn, C.~Durand, S.~Cheng, Y.~Chen, I.~Pasmans,
  and A.~Carrassi, 2024: {Accurate deep learning-based filtering for chaotic
  dynamics by identifying instabilities without an ensemble}. {\it Chaos\/},
  {\bf 34}, 091104.

\bibitem[{Bocquet and Sakov(2013)}]{boc13a}
Bocquet, M. and P.~Sakov, 2013: Joint state and parameter estimation with an
  iterative ensemble {K}alman smoother. {\it Nonlinear Proc. Geoph.\/}, {\bf
  20}, 803--818.

\bibitem[{Ginelli et~al.(2013)Ginelli, Chat\'{e}, Livi, and Politi}]{gin13a}
Ginelli, F., H.~Chat\'{e}, R.~Livi, and A.~Politi, 2013: Covariant {L}yapunov
  vectors. {\it J. Phys. A-Math. Theor.\/}, {\bf 46}, 254005.

\bibitem[{Lorenz and Emanuel(1998)}]{lor98a}
Lorenz, E.~N. and K.~A. Emanuel, 1998: Optimal sites for suplementary weather
  observations: simulation with a small model. {\it J. Atmos. Sci.\/}, {\bf
  55}, 399--414.

\bibitem[{Whitaker and Hamill(2002)}]{whi02a}
Whitaker, J.~S. and T.~M. Hamill, 2002: Ensemble data assimilation without
  perturbed observations. {\it Mon. Wea. Rev.\/}, {\bf 130}, 1913--1924.

\end{thebibliography}

\end{document}